\documentclass[final,1p,times]{elsarticle}

\usepackage{amssymb}
\usepackage{amsmath}

\journal{Physica A}

\begin{document}

\begin{frontmatter}



\title{Inverse Transitions in the Ghatak-Sherrington model with Bimodal Random Fields}


\author[ufpel,furg]{C.~V. Morais}\ead{carlosavjr@gmail.com}
\author[furg]{M. J. Lazo}
\author[ufsm]{ F. M. Zimmer}
\author[iff] {S.~G. Magalh\~aes}

\address[ufpel]{Instituto de F\'{\i}sica e Matem\'atica, Universidade Federal de Pelotas Caixa Postal 354, CEP 96010-090, Pelotas, Rio Grande do Sul, Brazil}

\address[furg]{Programa de P\'os-Gradua\c{c}\~{a}o em F\'{\i}sica - Instituto de Matem\'atica, Estat\'{\i}stica e F\'{\i}sica, 
Universidade Federal do Rio Grande, 
96.201-900, Rio Grande, RS, Brazil }

\address[ufsm]{Departamento de Fisica, Universidade Federal de Santa Maria,
97105-900 Santa Maria, RS, Brazil}

\address[iff]{Instituto de Fisica, Universidade Federal Fluminense, 24210-346
Niter\'oi, RJ, Brazil}

\begin{abstract}

The present work studies the Ghatak-Sherrington (GS) model with the presence of a longitudinal
magnetic random field (RF) $h_{i}$ following a bimodal distribution. The model considers a random
bond interaction $J_{i,j}$ which follows a Gaussian distribution with mean $J_0/N$ and variance
$J^2/N$. This allows us to introduce the bond disorder strength parameter $J/J_0$ to probe the combined effects of disorder coming from the random
bond and the discrete RF over  unusual phase transitions known as inverse transitions (ITs).
The results within a mean field approximation indicate that these two types of disorder have
complete distinct roles for the ITs.
They indicate that bond disorder creates the necessary conditions for the presence of inverse freezing or even inverse melting
depending on the bond disorder strength, while the RF tends to enforce mechanisms that destroy
the ITs. 
\end{abstract}


\end{frontmatter}
\section{Introduction}

Inverse Transitions (ITs) are counterintuitive reversible transitions, where the phase usually considered the more ordered has more entropy than the disordered one. 
ITs have been observed experimentally in very different physical systems, ranging from gold nanoclusters to high-Tc superconductors \cite{0,1,2,3,4,5,6,7,8}. 
Such diversity of physical systems displaying ITs have also triggered intense theoretical interest in which several models have been studied, in particular, magnetic ones because of its simplicity \cite{9,10,11,12,13,14,15,16,17}. From those magnetic models, it is suggested that disorder and frustration would be key ingredients to produce ITs, freezing (IF) or even melting (IM) \cite{17}. However, most of the previously mentioned studies
using magnetic models suffer from a limitation. The disorder comes basically from random bonds (RB) between spins. Therefore, it is an open issue whether or not other forms of disorder can contribute to produce ITs.

The family of Blume-Emery-Griffiths model \cite{BEG} as the Blume-Capel (BC) \cite{17a} and Ghatak-Sherrington (GS) ones \cite{18} have been quite useful to study the ITs. Therefore, they can be also helpful to answer the issue mentioned previously. These models consider bonds of spins 1 ($S=0, \pm 1$) and a crystal lattice field $D$ with phase diagrams exhibiting a tricritical point.
For instance, the GS model with a magnetic random field (RF) following a gaussian distribution has been recently applied to study the effects of the RF on ITs \cite{16a}.
This model is well known to present a reentrant first-order  phase transition between the spin glass (SG) phase and the paramagnetic (PM) one which characterizes an IF. 
The presence of such RF weakens the PM/SG first-order transition by lowering the tricritical point \cite{bcrf,bcrfpl}. 
Most important, it tends to suppress the IF.  
However, such analysis has been done in quite specific conditions:
in the limit of strong bond disorder and for a continuous distribution of the RF.  
Thus, a question remains open: are those effects found previously  still present
for other regimes than the strong bond disorder and for a discrete
distribution of the RF? For both cases one can expect that the phase diagram can become richer
with the presence of others thermodynamic phases as well as more complex boundary phases \cite{27}.

For this purpose, we studied here the GS model with the presence of a longitudinal magnetic RF $h_{i}$ following a bimodal distribution within weak disordered regimes for both the RF and the random bonds.
Therefore, the model considers a random bond interaction $J_{i,j}$ and the RF $h_i$. The interaction $J_{i,j}$ follows a Gaussian distribution with mean $J_0/N$ and variance $J^2/N$, which allows one to introduce the bond disorder strength parameter $J/J_0$ in the problem \cite{17}. 
The disorders are then treated by using the replica method, in which the free energy is obtained within the one-step replica symmetry breaking (1S-RSB) approximation \cite{Parisi}. 
In the present case, by adjusting the disorder strength $J/J_0$ parameter, 
the strong ($J/J_0>>1$) and non-disordered ($J/J_0=0$) regimes can be recovered. 
Most important, the study of  $J/J_0$  values corresponding to the weak disordered interactions 
could bring new information concerning mechanisms responsible for the ITs. 

It should also be remarked that the GS model presents the order parameter $r$ (see Eqs. \ref{eqnova2} and \ref{eqnova4}) related to the diagonal replica elements ($r=\langle S^{\alpha }S^{\alpha }\rangle$). This order parameter can
be used to obtain the average number of non-magnetic states per site $n_0=1-r$, which also reflects the capability of the sites to interact or not \cite{Castillo}. For instance, the limits $n_0=0$ and $n_0=1$ can be related to the fully interacting ($r=1$ high number of sites with spins $S=\pm1$) and the fully no interacting ($r=0$ high number of sites with $S=0$) regimes, respectively. 
As a consequence, there are two types of PM phases 
with  distinct $n_0$. This is an important condition for the  existence of the IF in the GS model \cite{10}. 

Moreover,  the  BC model with 
RF following a bimodal distribution is particularly interesting \cite{bcrf}. 
This model does not  spontaneously
display any IT at mean field level. 
However, it can also present the two types of PM
phases with distinct $n_0$. 
It means that one can have a PM phase with low entropic content. 
This fact can be explored to produce an IM  
in the BC model when an artificial entropic advantage of the interacting states is introduced \cite{9}. 
The IM is characterized by a transition from the PM phase at low temperature to the ordered ferromagnetic (F) phase at higher temperatures. 
Nevertheless, the bimodal RF is able to suppress the low entropic PM phase \cite{bcrf}, which could break an essential condition to the occurrence of IT \cite{11} in this specific model.  

On the other hand, 
the regime of weak RB in the GS model also introduces a F phase. 
However, in this is still not completely obvious if an IT can be observed. Thus, one can ask if the weak RB regime could be able to produce the necessary entropic advantage of the ordered phase to present spontaneously an IT.
Particularly this point has
not been yet studied in the context of  the IT problem. 
In addition, the effects of the bimodal RF could be more evident in the region of weak RB disorder. As a consequence, the superposition of the RB (already studied only for strong regimes) and discrete distributed RF (well studied in the BC model) effects on
the ITs could be more carefully analyzed. To accomplish this task, the effects of both RB and RF in the GS model can be traced from the behavior of 
$n_0$ \cite{18} to the entropic content of thermodynamic phases present in the problem. 
A detailed study of effects caused by different disorders on 
$n_0$ can be important to clarify the mechanism behind the existence of the IT in the present approach.

The paper is structured as follows: in Sec. II, the free energy within the 1S-RSB scheme is found.
In Sec. III, a detailed discussion of phase diagrams is presented. The last section is reserved for the
conclusions.

\section{General Formulation}

The model is given by the Hamiltonian
\begin{equation}
H = - \sum_{(i,j)} J_{ij} S_{i} S_{j} + D \sum_{i = 1 }^{N} S_{i}^2 - \sum_{i=1}^{N} h_{i}
S_{i},
\label{eq1}
\end{equation}
where the spin variables assume the values $S=\pm 1,0$, the summation $(i,j)$ is over any pair of
spins and $D$ is the crystal field. The spin-spin coupling $J_{ij}$ and the magnetic fields $h_i$ are
random variables following independent distributions

\begin{equation}
P(J_{ij})=\left[\frac{N}{2\pi J^{2}}\right]^{1/2} \exp\left[-\frac{N}{2J^{2}}\left(J_{ij} - J_0/
N\right)^{2}\right]
\label{eq2a}
\end{equation}
\begin{equation}
{\cal P}(h_i)=p\delta(h_i-h_0)+(1-p)\delta(h_i+h_0).
\label{eq2}
\end{equation}
Herein the procedure introduced in Refs. \cite{30,31} is closely followed. Therefore, the average
free energy per spin
\begin{equation} f=-1/(\beta N)\int dh_idJ_{ij}{\cal P }(h_i)P(J_{ij}) \ln Tr \exp{[-\beta
H(J_{ij},h_i)]}\end{equation}
can be obtained using the replica method as:
\begin{equation}
\label{eq5}
\begin{split}
\beta f
&= \lim_{n\rightarrow 0} \frac{1}{n} \left\lbrace
\frac{(\beta J)^2}{2} \sum_{\alpha<\gamma} q^{2}_{\alpha\gamma} + \frac{(\beta J)^2}{4}
\sum_{\alpha} q^{2}_{\alpha\alpha}\right.
\\& \left.
+\frac{\beta J_0}{2}\sum_{\alpha}m^2_{\alpha} - p\ln \mbox{Tr} e^{L^+} - (1-p) \ln \mbox{Tr}
e^{L^-}
\right\rbrace,
\end{split}
\end{equation}
where $\beta=1/T$ (T is the temperature), $\alpha$ and $\gamma$ are the replica labels, and
$L^{\pm}$ are the effective Hamiltonians given by
\begin{equation}
\label{eq7}
\begin{split}
L^{\pm} &= (\beta J)^{2} \sum_{\alpha<\gamma} q_{\alpha\gamma} S_{\alpha}S_{\gamma} +
\frac{(\beta J)^2}{2} \sum_{\alpha} q_{\alpha\alpha}S_{\alpha}^2 \\&+ \beta J_0 \sum_{\alpha}
m_{\alpha}S_{\alpha} -\beta D \sum_{\alpha} S_{\alpha}^2 \pm \beta h_0 \sum_{\alpha}
S_{\alpha}.
\end{split}
\end{equation}
In Eq. \ref{eq5} is used the steepest descent method (for $N\rightarrow \infty$) to evaluate the
fields $\{m\}$ and $\{q\}$ that are associated with the following correlations
\begin{equation}
\begin{split}
& m_\alpha=p<S_{\alpha}>_+ + (1-p)<S_{\alpha}>_- \\
& q_{\alpha\alpha}=p<S_{\alpha}^2>_++(1-p)<S_{\alpha}^2>_-,\\
& q_{\alpha\beta}=p<S_{\alpha} S_{\beta}>_++(1-p)<S_{\alpha} S_{\beta}>_-.
\label{eqnova2}
\end{split}
\end{equation}
The symbol $<...>_{\pm}$ means the thermodynamical average over the effective Hamiltonian
$L^{\pm}$ given in Eq. \ref{eq7}.

The Parisi's 1S-RSB is used to parameterize the replica matrix $\{q\}$ as shown below \cite{Parisi}
\begin{equation}
q_{\alpha\beta}=
q_{1} \mbox{ if } I(\alpha/x)=I(\beta/x)
\label{eq8.1}
\end{equation}
\begin{equation}
q_{\alpha\beta}=q_{0} \mbox{ if } I(\alpha/x)\neq I(\beta/x)
\label{eq8.2}
\end{equation}
where $I(\gamma)$ gives the smallest integer which is greater than or equal to $\gamma$. The replica diagonal
elements and the magnetization are respectively given by
\begin{equation}
q_{\alpha\alpha}=r~\mbox{and} ~m_{\alpha}=m .
\label{eqnova4}
\end{equation}

The parameterization introduced in Eqs. \ref{eq8.1}-\ref{eqnova4} is used in order the sum over
the replica labels in Eqs. \ref{eq5} and \ref{eq7}. This procedure results in quadratic terms that are
linearised introducing new auxiliary fields $z$ and $v$. Finally, the free energy is given by
\begin{equation}
\begin{split}
\beta f&= \frac{(\beta J)^{2}}{4}[(x-1)q_{1}^{2} - x q_{0}^{2} +r^{2}] + \frac{\beta J_0}{2}
m^2
\\ & - \frac{p}{x} \int {\cal D} z \ \ln\int {\cal D} v \ [ K^{+}(z,v)]^x \\
\\ & - \frac{1-p}{x} \int {\cal D} z \ \ln\int {\cal D} v \ [ K^{-}(z,v)]^x~,
\label{eq1aaasd}
\end{split}
\end{equation}
where
\begin{equation}
K^{\pm}(z,v)= 1 + 2 e^{\gamma} \cosh h^{\pm}(z,v)
\end{equation}
with $\int D w=\int\frac{d w}{\sqrt{2\pi}}e^{-w^{2}/2}$ $(w=z,v)$,
\begin{equation}
\gamma = \frac{(\beta J)^2}{2} (r - q_1) - \beta D,
\end{equation}
and $h^{\pm}(z,v)$ is
\begin{equation}
\begin{split}
h^{\pm}(z,v)&=\beta J[ \sqrt{q_0}z + \sqrt{ q_1-q_0}v] + \beta (J_0 m \pm h_0).
\end{split}
\label{e30sgs}
\end{equation}

The equations for the order parameters $q_{0}$, $q_{1}$, $r$, $m$ and the block size
parameter $x$ are obtained from Eq. \ref{eq1aaasd} by using the saddle point conditions. Other
thermodynamic quantities can also be derived from Eq. \ref{eq1aaasd}, as for instance, the entropy
$s=-\partial f/\partial T$. Particularly the present work analysis the symmetric case for the RF distribution in which $p=0.5$.

\section{Results}

This section displays phase diagrams $T/J_0$ versus $D/J_0$ that are build from numerical
solutions of the order parameters for several set of disorders: $J/J_0$ and $h_0/J_0$. In these, the
SG phase is characterized by RSB solution ($\delta\equiv q_1-q_0>0$) with zero magnetization
($m=0$). The ferromagnetic (F) phase occurs when the RS solution ($q=q_1=q_0$) is found with
$q> 0$ and $m>0$. The mixed ferromagnetic~$({\rm F^{\prime}})$ phase is given when $m\neq
0$ and $\delta >0$. The paramagnetic (PM) phase is characterized by $m=0$ with RS solution.
Particularly the RF induces the correlation given by $q$ in the PM phase, but without breaks the
replica symmetry ($\delta=0$, $q>0$ with $m=0$). The behavior of the entropy $s$ and $n_0$
\cite{Castillo} are also discussed in this section.

\begin{figure*}[t]
\includegraphics[width=9.5cm,angle=-90]{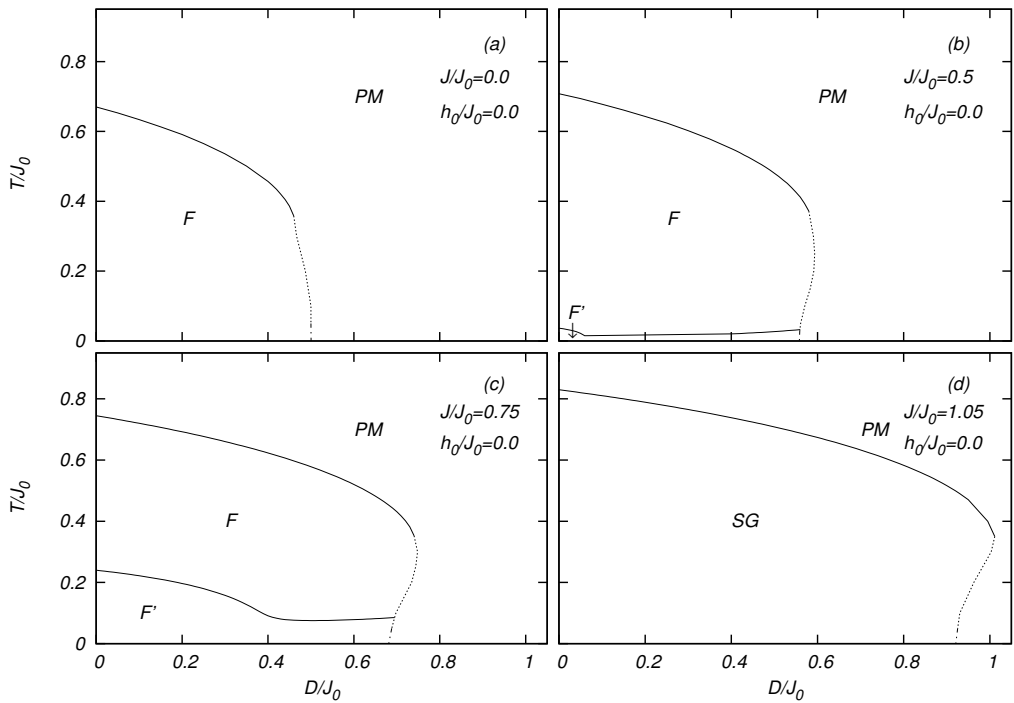}
\caption{Phase Diagrams $D/J_0$ versus $T/J_0$ for $h_0/J_0=0.0$ and several values of $J/J_0:
0.00, 0.5, 0.75$ and $1.05$. The full and the dotted lines represent the second and the first-order
transitions, respectively. The dashed-dotted lines at low T are extrapolations to the $T = 0 $.}
\label{fig1}
\end{figure*}

\begin{figure*}[ht]
\begin{center}
\includegraphics[height=6cm,width=5cm, angle=-90]{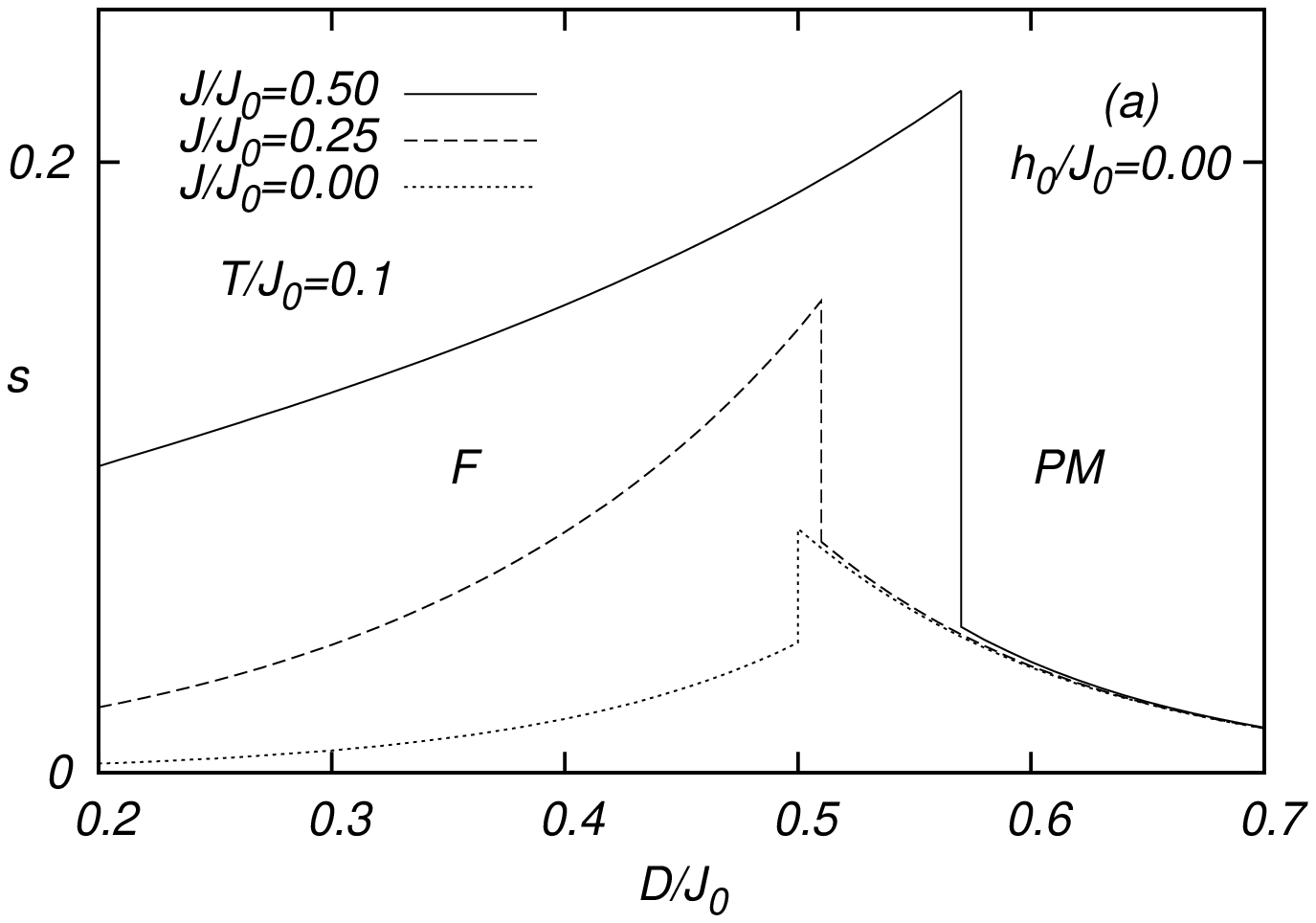}
\includegraphics[height=6cm,width=5cm, angle=-90]{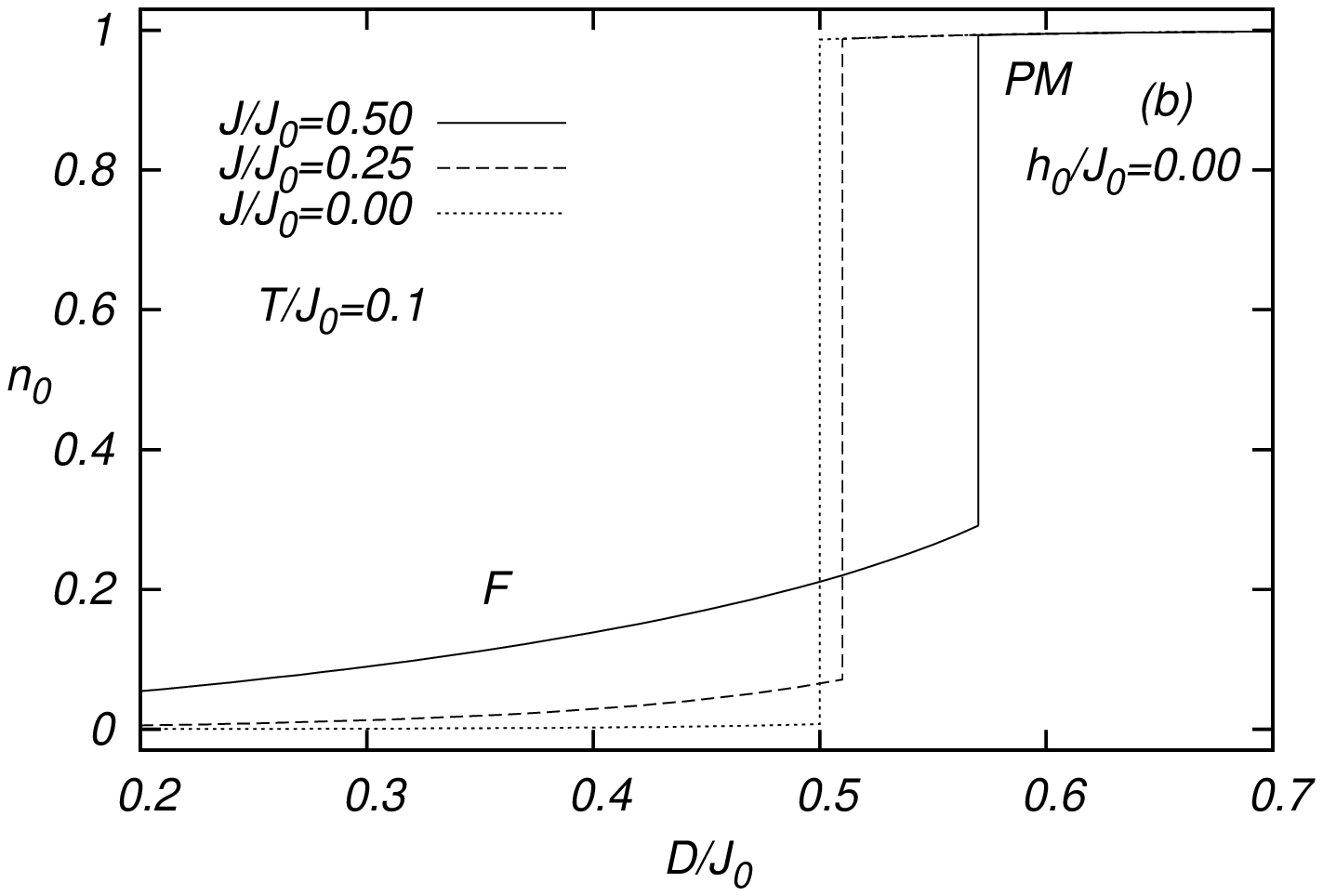}
\caption{ Behavior of (a) entropy $s$ and (b) $n_0$ as a function of $D/J_0$ for several values of
the strength of bond disorder $J/J_0$ without the presence of RFs.}
\label{fig2}
\end{center}
\end{figure*}

\begin{figure*}[ht]
\begin{center}
\includegraphics[width=5.5cm,angle=-90]{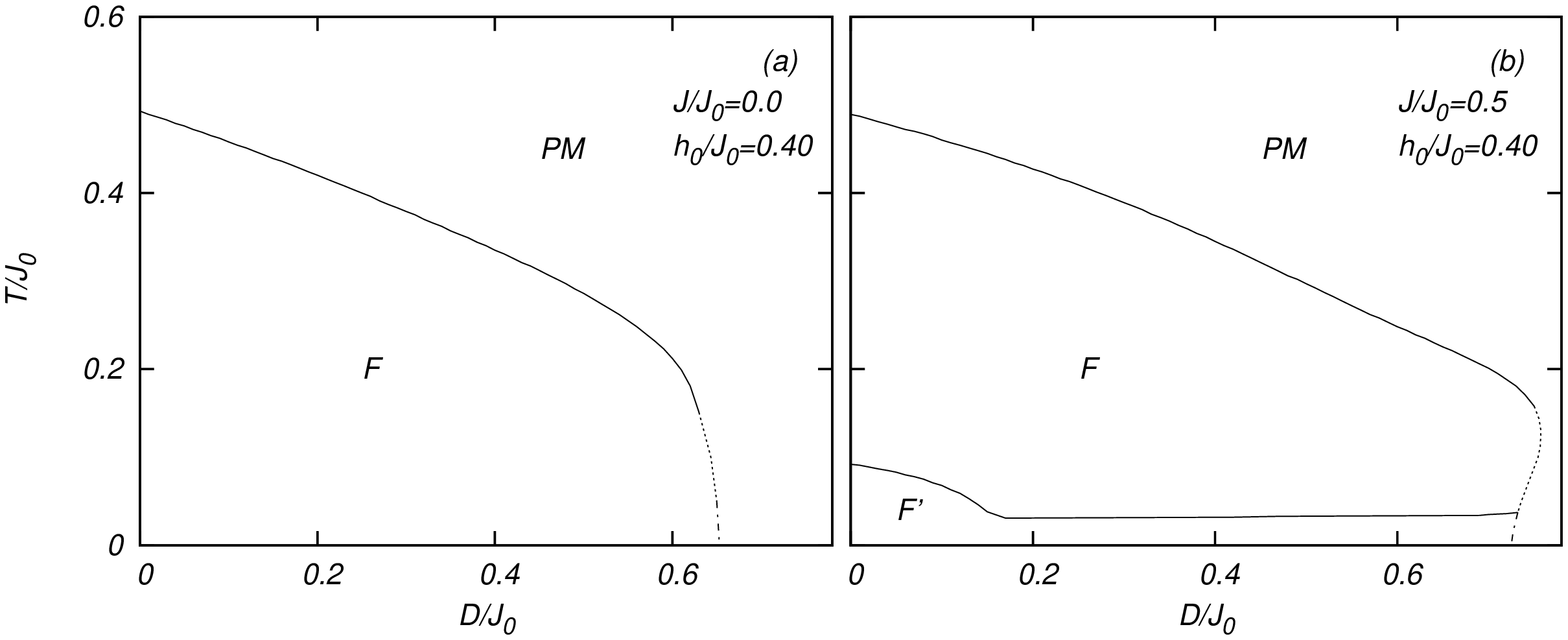}
\caption{Phase diagrams $D/J_0$ versus $T/J_0$ for $h_0/J_0=0.4$ and $J/J_0:0.0,0.5$. The full
and the dotted lines represent the second and the first-order transitions, respectively. The dashed-
dotted lines at low T are extrapolations to the $T = 0$. }
\label{fig3}
\end{center}
\end{figure*}
The phase diagrams exhibited in Fig. \ref{fig1} consider $h_0/J=0.0$ with several values of the
strength of bond disorder $J/J_0: 0.0,0.5,0.75$ and $1.05$. For example, Fig. \ref{fig1}(a) shows
the phase diagram without any type of disorder ($h_0/J_0=0.0$ and $J/J_0=0.0$). In this case,
the continuous PM-F phase transition is decreased until a tricritical point as $D/J_0$ increases.
After the tricritical point, the phase transition becomes first-order where there are more than one
PM and F solutions. The first-order boundary is then located by comparing the lower free energy
of the PM and F (or SG) solutions by using the same criteria proposed in Ref. \cite{salinas}. 
One important point here is that the 
1S-RSB Ansatz is still unstable against fluctuations in the replica
space and the right thermodynamics of the SG solutions can be obtained only with full 
replica symmetry break method. However, the 1S-RSB procedure can give reliable results
around the phase transition (including the reentrance) \cite{9,10}. 
Particularly, the free energy with $J/J_0=0$ and $h_0/J_0=0$ (Eq.
\ref{eq1aaasd}) presents the same equation as that one obtained in Ref. \cite{9} for the Blume-
Capel model. In fact, the respective phase diagrams found in Refs. \cite{9,BEG} are recovered in Fig.
\ref{fig1}(a) where there is no reentrant behavior. However, a very weak strength of bond disorder
($J/J_0$) affects strongly the first-order PM-F transition in such way that a reentrant behavior
appears. For instance, Fig. \ref{fig1}(b) displays a reentrance in the first-order PM-F transition
that seems to be intensified as $J/J_0$ increases in Fig. \ref{fig1}(c). Such transition represents
an IM, in which the usual entropic relation between the ordered F phase and the 
PM
phase is changed (see Fig. \ref{fig2}(a)). In addition, Figs. \ref{fig1}(b) and \ref{fig1}(c) also
indicate the presence of a mixed frustrated $F'$ phase for low $T/J$, which is obtained within the
1S-RSB procedure. On the other hand, for the strong disordered regime, the F phase is replaced
by the SG one (see Fig. \ref{fig1}(d) for $J/J_0=1.05$). In this case there is a continuous PM-
SG phase transition that decreases until a tricritical point $T_{tc}$ with increasing of $D/J_0$.
Below $T_{tc}$, a reentrant first-order PM-SG phase transition is obtained which indicates an IF.
Therefore the intensity $J/J_0$ plays an important role in the first-order reentrant
transitions. Most importantly, the results presented in Fig. \ref{fig1} suggest that the present model
is able to show both the IF and the IM. The precise type of IT depends on the strength of the bond
disorder.

The mechanism underlying these reentrances can be explained analyzing the entropy $s$ and the
average number of the non-magnetic sites $n_0$. For example, Fig. \ref{fig2}(a) exhibits $s$ vs
$D/J_0$ for $h_0/J_0=0$ and several values of $J/J_0$ at low temperature ($T/J_0=0.1$). For $J/
J_0=0$ (case without reentrance), the entropy of the ordered F phase is smaller than the PM one at
the first-order transition. However, the entropy of the ordered phase increases with $J/J_0$. In fact,
the entropy of the PM phase is practically unaffected by increasing $J/J_0$, while the entropy of
the ordered phase suffers a large increment (see Fig. \ref{fig2}(a) for $J/J_0>0$). Fig. \ref{fig2}
(b) shows the behavior of $n_0$ for the same values of strength of disorder and temperature as Fig.
\ref{fig2}(a). This result could elucidate why only the entropy of the F phase is modified when the
$J/J_0$ is adjusted. Basically, the results indicate that $n_0$ of the F phase is increased by $J/J_0$.
However, $n_0$ remains unaltered in the PM phase that also presents the higher values of $n_0$
(high number of sites with spin $S=0$). In other words the low temperature PM phase presents few
spin configurations available when $n_0\rightarrow 1$. As a consequence the entropy of this PM
phase is very low even for higher $J/J_0$ values. It is in contrast with the high temperature PM
phase that shows a strong thermal spin fluctuation.

\begin{figure*}[th]
\includegraphics[width=6cm,width=5cm, angle=-90]{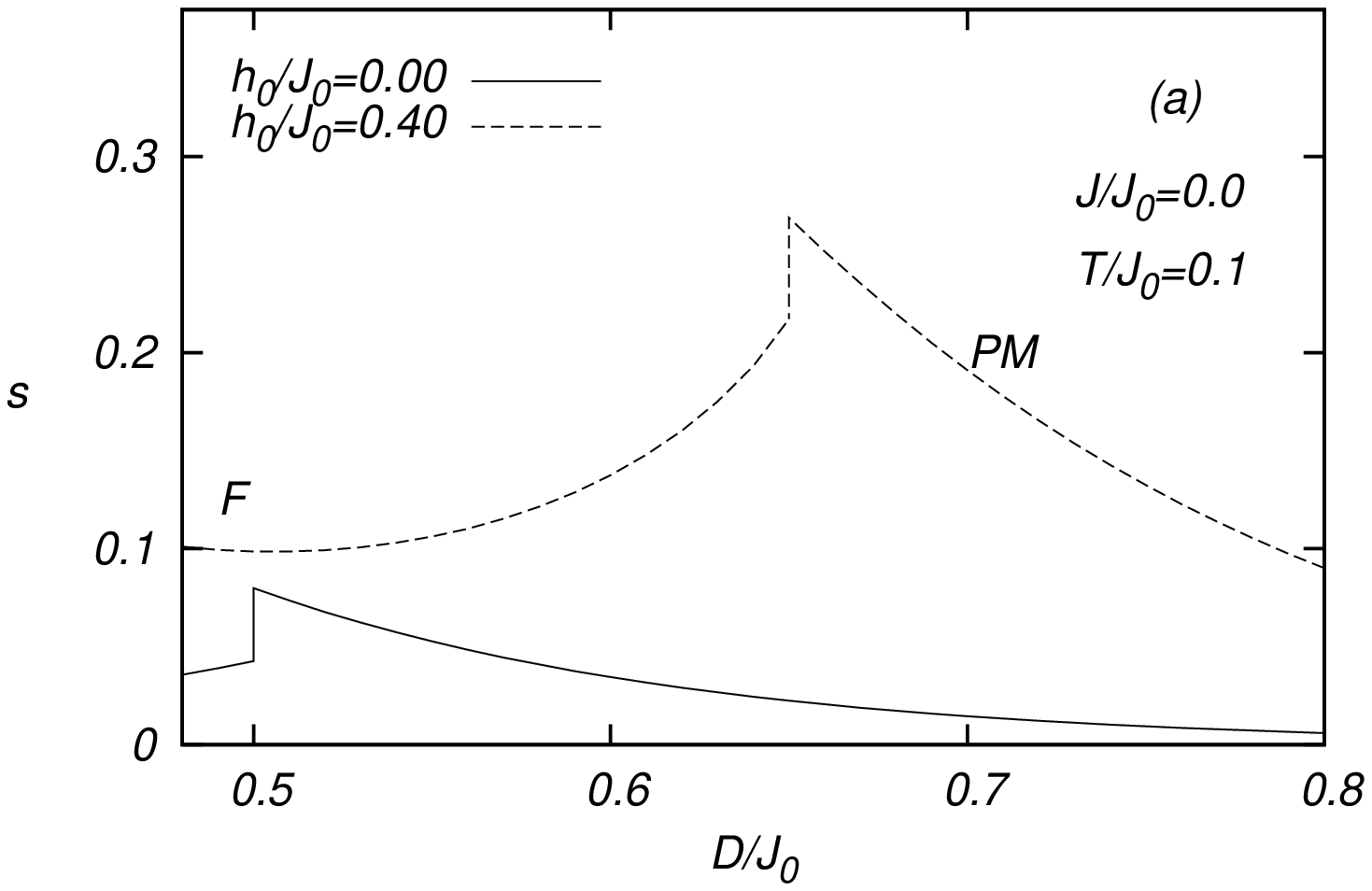}
\includegraphics[width=6cm,width=5cm, angle=-90]{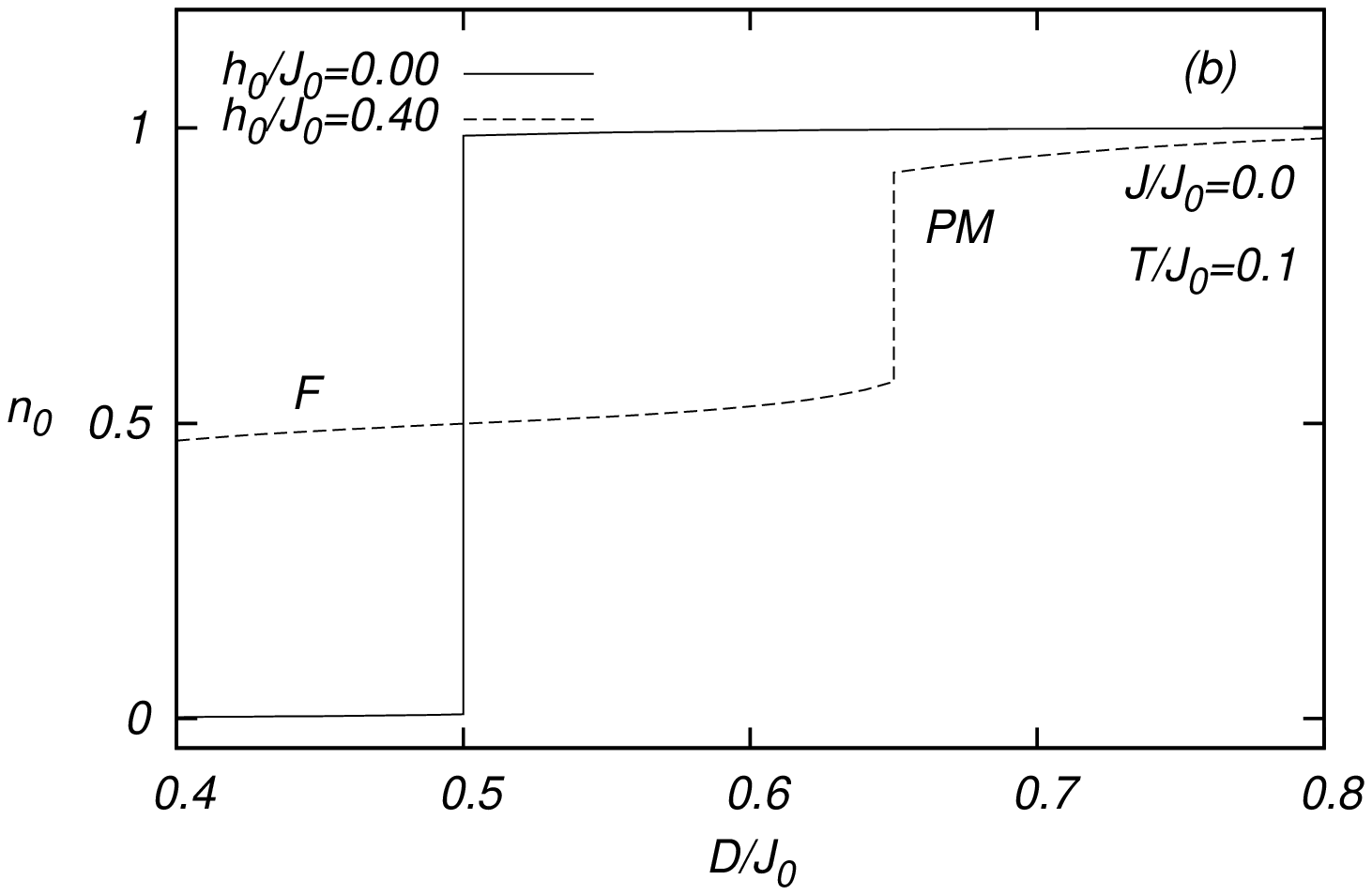}
\caption{Behavior of (a) entropy $s$ and (b) $n_0$ as a function of $D/J_0$ for $T/J_0=0.1$,  two values of the
strength of bond disorder $J/J_0$ and two values of $h_0/J_0:0.0,0.4$. }
\label{fig4}
\end{figure*}

On the other hand, the bimodal RF causes different effects than those ones caused by $J/J_0$.
For instance, Fig. \ref{fig3}(a) shows the phase boundary for $h_0/J_0=0.4$ and $J/J_0=0$,
where the only effect of disorder comes from the bimodal RF. It is important to remark that for $J/
J_0=0$, a coexistence of ordered FM solutions can appear depending on the value of the RF
\cite{bcrf}. However, for the purposes of the present work, only the PM/FM border and the effects
of the $h_0/J_0$ on it are analyzed. In this case, the critical temperature and the tricritical point
are displaced for lower values of $T/J_0$ and higher values of $D/J_0$ when $h_0$ increases
(compare Fig. \ref{fig1}(a) with Fig. \ref{fig3}(a)). Most important, the presence of $h_0$ is
unable to produce any reentrant behavior. Fig. \ref{fig3}(b) allows one to analyze the effects caused
by the superposition of both disorders: $h_0/J_0=0.4$ and $J/J_0=0.5$. It exhibits a reentrance
(IM) attributed to the increase of the random bond disorder that also introduces a mixed phase at
very low temperatures.

The analyses of the entropy and $n_0$ in the presence of RF with $J/J_0=0$ are accomplished
in Fig \ref{fig4}. Comparing the cases $h_0/J_0=0.0$ and $h_0/J_0=0.4$ at low temperature $(T/J_0=0.1)$ in Fig. \ref{fig4}(a), it
is possible to observe that not only the entropy of the F phase increases with $h_0/J_0$, but also
the entropy of the PM one. It is important to observe that the entropy of PM phase depends on
the strength of the RF disorder. This increase in $s$ in the PM phase can be one of the reasons for the
incapability of the RF disorders generate IT. The behavior of $n_0$ in Fig. \ref{fig4}(b) can help
us to understand better this point. For instance, the $n_0$ is changed by the strength of RF in both
phases the ordered and the disordered. In particular, the PM phase at lower temperature presents
a diminishment in $n_0$ as $h_0/J_0$ increases. This is an important difference in relation to
the random bond disorder. As a consequence the RF causes an increase in the entropy of the PM
phase.


\section{Conclusion}


In the present work, the Ghatak-Sherrington SG model has been studied with the addition of
longitudinal magnetic random field (RF) $h_{i}$. This study was carried out within the 1S-RSB
scheme using a mean field approximation. This approach can be helpful to elucidate the role of
different kind of disorders (RB
and RF) in the so called Inverse Transitions, melting or freezing.

According to our results, the crystal field $D$ favors energetically the spin states $S=0$, that favoring introduces a PM phase for larger $D$. 
However, for no disorder ($J/J_0=0$ and $h_0/J_0=0$), the PM entropic content is always larger than that one of the F phase. However,
the situation changes at very lower temperature when the random bond disorder is considered
($J/J_0>0$). For a small strength of bond disorder ($0<J/J_0<1$) the F phase is still found, but
now with larger entropy as compared with the $J/J_{0}=0$ case, while the PM entropy is almost
unchanged. In other words, it arises an IM for a certain range of $D$. As the strength of the random bond increases ($J/J_0>1$) the SG phase replaces the F order and the IF occurs. On the other hand,
the presence of a RF  increases the entropic content not only of the ordered F or
SG phases, but also of the disordered PM phase. It is important to remark this particular point that
the RF causes a strong increase of the PM phase entropy. Therefore, it breaks one of the essential
conditions to the occurrence of IT that is the existence of a PM phase with lower entropy than the
ordered phase. In fact, the RF tends to couple with the interacting spin states ($S=\pm 1$) that can
be favored energetically even for higher values of $D$. As a consequence, the PM phase at lower temperature with smaller entropy is hard to be found.

To summarize the main difference between the two kinds of disorder studied here in the IT context
is the effect caused in the PM phase. The RFs compete with $D$ by increasing the entropy of the
PM phase of lower temperatures, while the random bonds affect only the entropic content of the
ordered phase favoring the IT.

\section*{Acknowledgments}
This work was partly supported by the Brazilian
agencies CNPq, CAPES, FAPERJ, and FAPERGS.

\end{document}